\documentclass[aps,pre,superscriptaddress,10pt,twocolumn,amsmath,floatfix]{revtex4}

\usepackage{graphicx}
\usepackage{amsmath, amsthm, amssymb}

\begin{document}
\title{Effects of quenched disorder on critical transitions in pattern-forming systems}
\author{Hezi Yizhaq}
\affiliation{Department of Solar Energy and Environmental Physics, Blaustein Institutes for Desert Research, Ben-Gurion University of the Negev, Sede Boqer Campus 84990, Israel}
\affiliation{The Dead Sea and Arava Science Center, Tamar Regional Council, Israel}
\author{Golan Bel}
\affiliation{Department of Solar Energy and Environmental Physics, Blaustein Institutes for Desert Research, Ben-Gurion University of the Negev, Sede Boqer Campus 84990, Israel}
\date{\today}

\begin{abstract}
Critical transitions are of great interest to scientists in many fields. Most knowledge about these transitions comes from systems exhibiting the multistability of spatially uniform states. In spatially extended and, particularly, in pattern-forming systems, there are many possible scenarios for transitions between alternative states. Quenched disorder may affect the dynamics, bifurcation diagrams and critical transitions in nonlinear systems. However, only a few studies have explored the effects of quenched disorder on pattern-forming systems, either experimentally or by using theoretical models. 
Here, we use a fundamental model describing pattern formation, the Swift-Hohenberg model and a well-explored mathematical model describing the dynamics of vegetation in drylands to study the effects of quenched disorder on critical transitions in pattern-forming systems.
We find that the disorder affects the patterns formed by introducing an interplay between the imposed pattern and the self-organized one.
We show that, in both systems considered here, the disorder significantly increases the durability of the patterned state and makes the transition between the patterned state and the uniform state more gradual. In addition, the disorder induces hysteresis in the response of the system to changes in the bifurcation parameter well before the critical transition occurs. We also show that the cross-correlation between the disordered parameter and the dynamical variable can serve as an early indicator for an imminent critical transition.
\end{abstract}

\maketitle
\section{Introduction}
Pattern formation is a spectacular phenomenon of self-organized heterogeneity in an otherwise spatially
homogeneous background. Spatial patterns can be found in many natural systems, including clouds~\cite{sengupta1990cumulus}, sand
ripples \cite{hansen2001pattern,yizhaq2012}, chemical reactions \cite{lee1993pattern}, fluid dynamics \cite{greenside1984pattern}, electromagnetic properties of materials \cite{Dagotto2005} and
vegetation in drylands \cite{Borgogno2009}. Mathematical models that explain the emergence of the patterns are based on the
positive and negative feedbacks that lead to the nonlinear dynamics of the systems \cite{hoyle2006pattern}. The pattern formation
introduces additional complexity into systems exhibiting critical transitions. Most of our knowledge about
pattern formation originates from controlled laboratory experiments \cite{Chandrasekhar1961,CrossHohenberg1993,Bod2000,Field1972,Field1974,Field1985}. Due to the
relevance of critical transitions to many systems affecting our lives \cite{Scheffer2001,scheffer2009critical}, a great deal of effort has been
made to identify early warning signals for critical transitions \cite{Tongway1995,Kefi2007,Scanlon2007,scheffer2009early,Scheffer2012}.
Although much progress has been made in our understanding of pattern formation in spatially extended
systems and its role in regime shifts \cite{sherratt2013,vanderstelt2013,zelnik2013,siteur2014}, there are still many open questions, such as the effects of additive and
multiplicative noise on the pattern-forming systems, the dynamics of fronts between different patterns, the
interplay between different pattern-forming mechanisms \cite{kinast2014} and the effects of localized states on the critical transitions (recently, it was suggested 
that the existence of stable localized states in some pattern-forming systems may affect their critical transitions, 
and a gradual regime shift may be observed, rather than the expected catastrophic shift \cite{bel2012gradual}). In addition, very little is known about the effects of quenched disorder (i.e., parameters defining the dynamics of the system which are random variables that do not evolve with time) on the dynamics, stability and durability of the pattern-forming system.
 
It has been suggested by several authors
that the disorder may play an important role in the dynamics of nonlinear systems and their bifurcation
diagrams \cite{scheffer2009critical,nystrom2001spatial,connell1983evidence,peterson2000scaling,petraitis1999importance,van2005implications,bonachela2015}.
For instance, it was shown that in a system exhibiting bistability of two uniform states, quenched disorder does not affect the response of the system to changes in the driving force \cite{van2005implications}.
The effects of spatially periodic heterogeneity and simple realizations of heterogeneity \cite{Voroney1996,bar2002,page2003,page2005,haudin2010,clerc2010} on the pattern formation and on the range of Turing instability have been studied, with results
showing that heterogeneity strongly affects the pattern selection and Turing space.
However, only a few studies have explored the effects of quenched disorder on pattern-forming systems,
either experimentally or by using theoretical models \cite{reichhardt2003,olson2004dynamics,carillo2004,delmaestro2006,getzin2006,getzin2008,lee2011,sheffer2013emerged}. 
The dynamics of stripes, formed by
particles with a long-range Coulomb repulsion and a short-range exponential attraction, was investigated in
\cite{reichhardt2003,olson2004dynamics}. The molecular dynamic simulations showed that there is a critical value of the disorder strength
above which there is no pattern formation. It was also shown that the effect of the disorder may be overcome
by introducing a DC field. Effects of disorder on pattern formation were also observed in a model for cuprate
superconductors \cite{delmaestro2006}. In \cite{lee2011}, a method for studying pattern formation in ion traps with controlled disorder
and noise was suggested. A more general study focused on the role of quenched disorder in the effect of
spatial coherence resonance \cite{carillo2004}.

Here, we use a simple model and a more complex one to study the effects of quenched disorder on critical transitions in pattern-forming systems. In Sec. \ref{sec:SH}, we investigate the Swift-Hohenberg model with quenched disorder in the coefficients of the positive and negative feedbacks. It is shown that the disorder significantly affects the critical transitions, and we also suggest an early indicator for an imminent transition. In order to verify that the results are not limited to the simple model and in order to demonstrate their importance in ecosystem dynamics, we use a well-explored model describing the dynamics of water-limited vegetation in Sec. \ref{sec:R}. The disorder in this model represents the landscape heterogeneity that is found in most ecosystems. In Sec. \ref{sec:disc}, we discuss the results and their implications for the specific systems we investigated, as well as for other pattern-forming systems. 

\section{The Swift-Hohenberg model}\label{sec:SH}
The Swift-Hohenberg model is often considered as the simplest model to describe the dynamics of a pattern-forming system.
It was first suggested in the context of convective instability \cite{SH1977} and has since become a prototype model of pattern formation.
The finite wavenumber instability arises due to the interplay between short-range positive feedback and long-range negative feedback.
The dimensionless form of the model equation is:
\begin{equation}\label{eq:SH}
 \frac{\partial u}{\partial t}=ru+bu^2-cu^3-(\nabla^2+q_c^2)^2u.
\end{equation}
The growth rate $r$ will be used as the bifurcation parameter describing the growth-limiting factor. The nonlinear terms represent positive and negative local feedbacks.
The negative diffusion term represents the short-range positive feedback, and the fourth order derivative represents the long-range negative feedback.
In the contexts of population and ecological systems dynamics, the positive feedbacks are often referred to as activation or facilitation while the negative feedbacks are referred to as inhibition or competition.
The Swift-Hohenberg model has been studied extensively in various contexts (e.g., \cite{Pomeau1986,Firth2007,Short2010,Mercader2011,Lloyd2013,Gandhi2015,Tripathi2015}). Here, we use this model to study the effects of quenched disorder on the dynamics of pattern-forming systems and, in particular, on the critical transitions in these systems.
We introduced disorder in the parameters $b$ and $c$ (we only considered one disordered parameter at a time) using a uniform distribution that is fully characterized by its width, $v$, and its mean.
The distribution of the disordered parameter is given by
\begin{eqnarray}\label{udist}
p_U(b) = \left\{ \begin{array}{l}
 \frac{1}{v}\quad \quad \left\langle {b} \right\rangle -\frac{v}{2} \le b  \le \left\langle {b} \right\rangle+\frac{v}{2} \\[5pt]
 0\quad \quad \quad \quad \quad \quad else \\
 \end{array} \right.,
\end{eqnarray}
where $b$ is the disordered parameter, $\langle b \rangle$ is its mean value and its variance, $\sigma^2=v^2/12$.
In this study, we kept constant the parameter $q_c=1$.
The simulations presented here were performed on a lattice with 1024 sites; the total length was set to $65\pi$, and periodic boundary conditions were applied.
Spatial derivatives were approximated using a second-order finite difference scheme, and the temporal integration was based on the first-order Adam-Bashforth method with $dt=10^{-4}$.
The convergence to a steady state of the solutions presented throughout the manuscript was verified by inspecting the maximal change in the value of the dynamical variables between $t=t_{max}$ and $t=t_{max}/2$.  

\begin{figure}[!ht]
   \centering
   \includegraphics[width=0.9\linewidth]{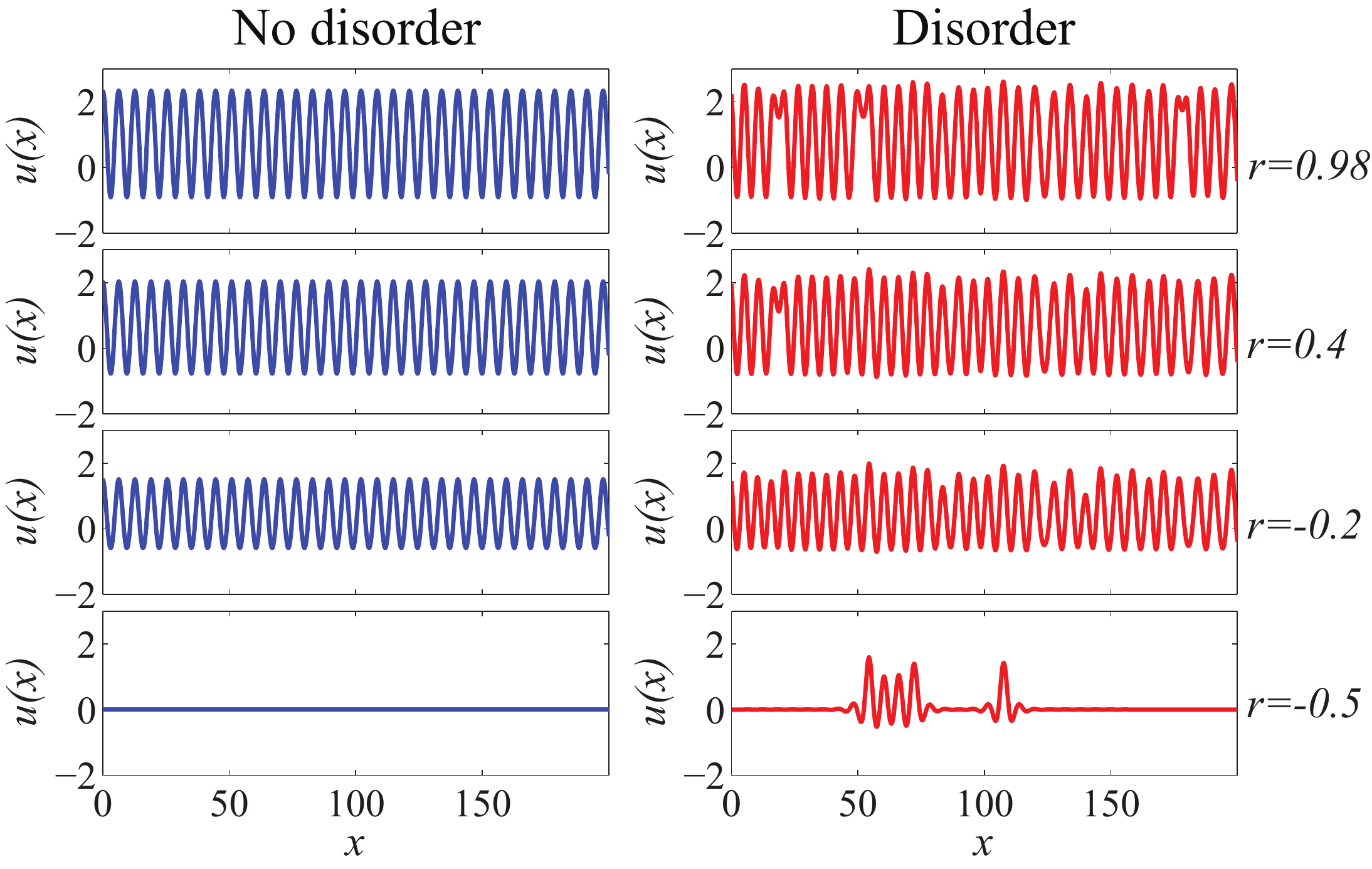}
   \caption{The effects of quenched disorder on the stable steady state patterns predicted by the Swift-Hohenberg model.
   The different rows correspond to $r=0.98,0.4,-0.2,-0.5$ as specified in the figure. The left column presents the patterns in the absence of disorder while the right column presents their counterparts in the presence of disorder in the parameter, $b$. The model parameters are: $c=1$ and $\langle b \rangle=1.8$. In the right panel, the width of the distribution is $v=2$. The disorder distorts the patterns and, in addition, enables non-zero values of $u$ under conditions in which the homogeneous system converges to the uniform $u=0$ state. }
   \label{fig:SHpatt}
 \end{figure}

\begin{figure*}[!ht]
   \centering
   \includegraphics[width=0.9\linewidth,height=0.3\linewidth]{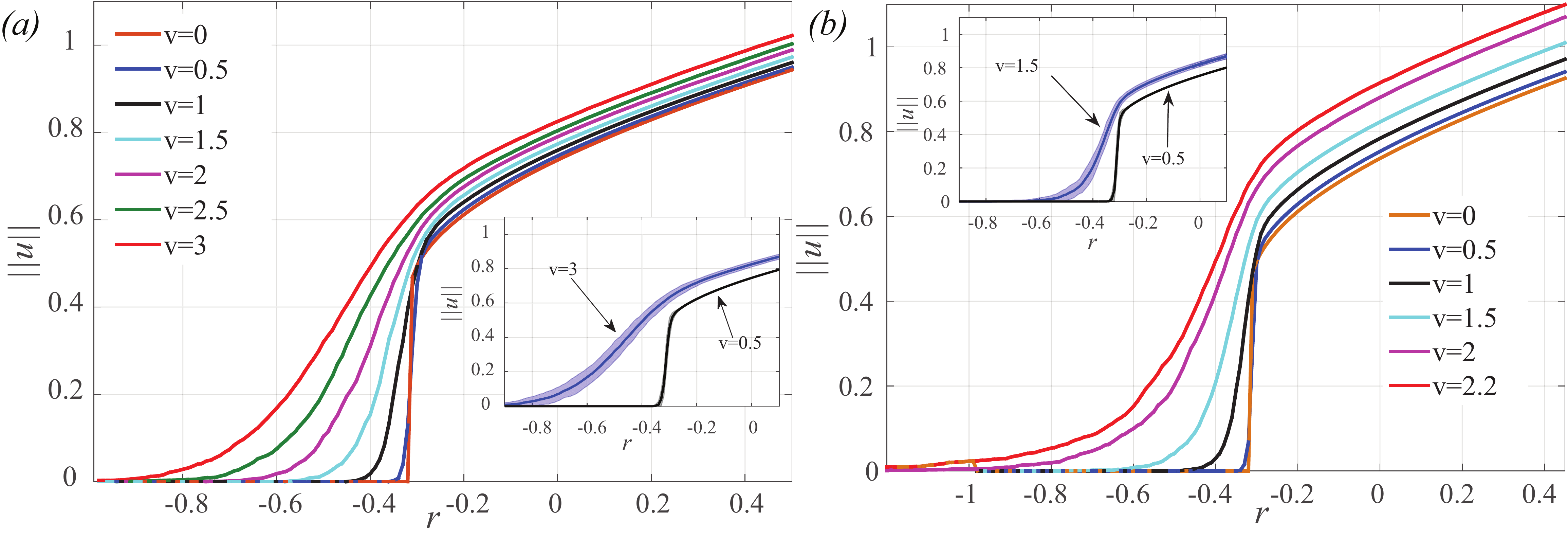}
   \caption{The effects of quenched disorder on the bifurcation diagram of the Swift-Hohenberg model.
   The left column (a) presents the effects of disorder in the parameter $b$ ($\langle b \rangle=1.8$ and $c=1.1$) while the right column (b) presents the effects of disorder in the parameter $c$ ($\langle c \rangle=1.1$ and $b=1.8$).
   The different lines correspond to different strengths of the disorder as specified in the legends. The insets show the two distinct strengths of the disorder. The shaded areas in the insets represent the uncertainty associated with each curve. The uncertainties were derived from the standard deviation of 50 realizations of the quenched disorder. It is apparent that disorder in any of these parameters extends the existence range of the patterned states. }
   \label{fig:SHbif}
 \end{figure*}

\begin{figure*}[!ht]
   \centering
   \includegraphics[width=0.9\linewidth]{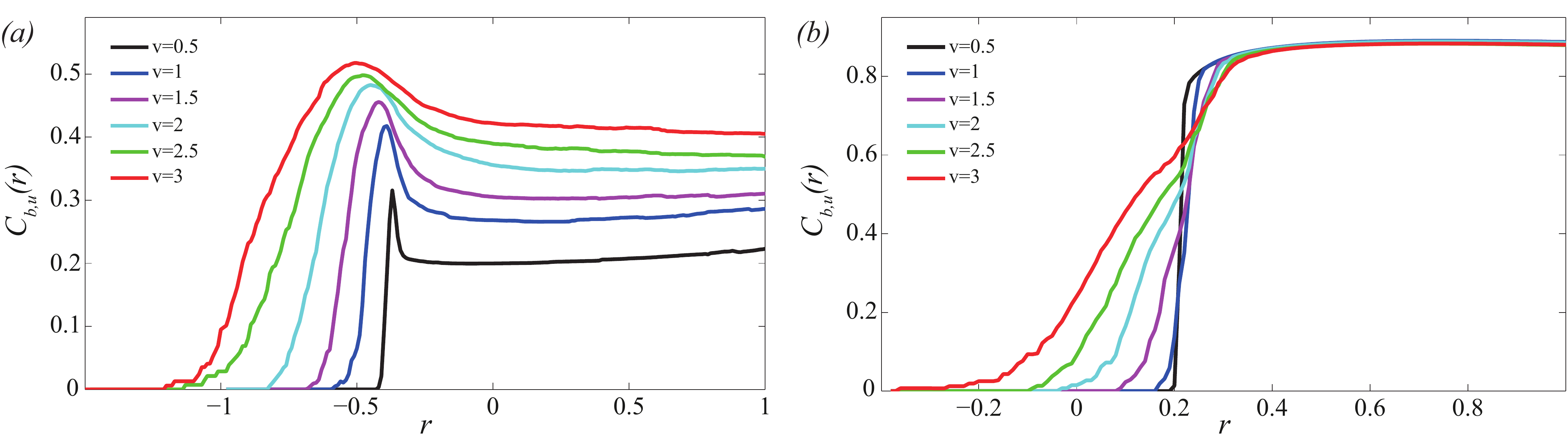}
   \caption{The cross-correlations between the disordered parameter and the dynamical variable in steady states of the Swift-Hohenberg model and a simplified model with only spatially uniform steady states.
   The left column (a) presents the cross-correlations for patterned steady states while the right panel (b) presents their counterparts in the modified model allowing only uniform steady states. In both models, the disorder was introduced in the parameter $b$ ($\langle b \rangle=1.8$ and $c=1$).
   The different lines correspond to different strengths of the disorder as specified in the legends. 
   In the model with only uniform steady states, the cross-correlation is always high, corresponding to the higher values of $u$ in locations with stronger positive feedback (i.e., larger $b$). On the contrary, in the pattern-forming system, the cross-correlation is lower due to the short-range facilitation effect that allows higher values of $u$ in domains with lower positive feedback. However, as the system approaches the critical transition, the cross-correlation increases significantly. This increase may serve as an early indicator for imminent critical transitions.}
   \label{fig:cross}
 \end{figure*}

We started the analysis by studying the effects of the disorder on the steady state patterns. For simplicity, we focused on one spatial dimension. Figure \ref{fig:SHpatt} presents the steady state patterns for different values of the bifurcation parameter, $r$ (the value corresponding to each row is specified in the figure). The left column corresponds to a homogeneous system, and the right column corresponds to a system with disorder in the amplitude of the local positive feedback, $b$. We used the uniform distribution (Eq. \eqref{udist}) with $\langle b \rangle=1.8$ and $v=2$. For high values of the bifurcation parameter, the disorder results in distorted patterns, as expected. For lower values (the lowest row of Fig. \ref{fig:SHpatt}) of $r$, the homogeneous system converges to the uniform zero state while the disordered system shows localized distorted patterns.
Our main interest is in the effects of the disorder on critical transitions. Therefore, we studied the effects of disorder in the parameters representing the strength of the positive and negative local feedbacks ($b$ and $c$, respectively). Figure \ref{fig:SHbif} depicts the L2-norm of the dynamical variable $u$ versus the bifurcation parameter, $r$, for different values of the disorder. The left panel (a) corresponds to disorder in $b$, and the right panel (b) corresponds to disorder in $c$. The different lines correspond to different strengths of the disorder (the strength of the disorder is quantified by the width of its distribution, $v$). Each line represents the average value of 50 realizations of the disorder. The insets in both panels show the curves for two distinct values of $v$ with the shaded area representing the standard deviation of the 50 realizations at each value of $r$. The insets demonstrate that the standard deviation is smaller than the difference between the different lines in the main panels. Both panels show that the stronger the disorder, the lower the value at which the system collapses to the uniform $u=0$ state. Moreover, the stronger the disorder, the more gradual the transition is from the patterned state to the zero uniform state.
To better understand the effects of the heterogeneous parameters on the dynamics and on the critical transition, we investigated the cross-correlations between the dynamical variable, $u$, and the disordered parameter (we present here the results for the disorder in $b$, but similar results were obtained for the disorder in $c$). Because the system includes nonlocal terms, the cross-correlation is maximal when considering the average value of $b$ in the vicinity of each point. The length over which $b$ should be averaged is determined by the diffusion coefficient. In the appendix \cite{SM}, we provide a more detailed analysis of the optimal averaging length.
Figure \ref{fig:cross} presents the cross-correlations between $u$ and $b$ versus the bifurcation parameter. The left panel (a) corresponds to the full Swift-Hohenberg model (Eq. \eqref{eq:SH}) and different strengths of the disorder. The cross-correlation is almost constant away from the critical point, and it increases significantly as the system approaches the critical point. After the system collapses to the zero uniform state, it reaches zero, reflecting the fact that $u=0$, regardless of the value of $b$. In the right panel (b), we present the same cross-correlation for a non-pattern forming system that was obtained by setting the coefficient of the fourth derivative to zero and changing the sign of the coefficient of the second derivative (to ensure convergence of the system) in Eq. \eqref{eq:SH}. This simplified system exhibits a bistability of two uniform states. The cross-correlation here is much higher than in the pattern-forming system, and it decreases to zero as the system undergoes the critical transition. No increase, or peak, is seen in this case.
To verify that the results presented above are not unique to the Swift-Hohenberg model and are applicable to other, more complex, pattern-forming systems, we consider, in the next section, a well-explored model describing the dynamics of dryland vegetation. Spatially extended ecosystems are often heterogeneous at various scales. Moreover, the environmental conditions vary due to climate dynamics and interactions with other ecosystems. Much is known about regime shifts (i.e., critical transitions) in homogeneous systems. The model studied in the next section will help us to shed light on the effects of the oft-neglected quenched disorder on the dynamics of pattern-forming ecosystems both close to and away from the critical points.

\section{The dynamics of water-limited vegetation}\label{sec:R}

To demonstrate the generality and the importance of the effects of quenched disorder on pattern-forming systems, we use the context of water-limited vegetation dynamics.
Vegetation patterns have been observed in many places
around the world, including Africa, Australia, North and South America, and Asia \cite{deblauwe2008global}.
Vegetation pattern formation is commonly seen as a self-organization phenomenon due to competition for water and positive feedbacks.
The positive feedbacks considered in different models include infiltration contrast, water uptake and root augmentation \cite{zelnik2013}. The interplay between these feedbacks is highly non-trivial and affects the patterns formed by the vegetation \cite{kinast2014}.
The dynamics of the vegetation and the water variables is described by a set of nonlinear reaction-diffusion equations.
These models predict five basic vegetation states that are observed along a decreasing rainfall gradient--uniform
vegetation cover, vegetation cover interspersed with gaps of bare soil, vegetation stripes, vegetation spots, and uniform
bare soil \cite{Borgogno2009,Jost2001,rietkerk2004selfa}--and the existence of a bistability range for each pair of consecutive states, e.g., 
bistability of stripes and spots. The extensive research of vegetation
models and the deep knowledge of these systems make them an ideal choice for the study of
the effects of quenched disorder. The backgrounds for the vegetation patterns are the landscapes in which they occur.
Almost all soil in nature is heterogeneous. This heterogeneity arises from small changes in the soil texture and
composition, the presence of small rocks and stones, micro-topography, changes in soil depth \cite{sela2012soil}, changes
in the slope, the spatial distribution of nutrients \cite{cambardella1994field,schlesinger1996spatial} and many other factors \cite{mulla1999}.
\begin{figure}[!ht]
   \centering
   \includegraphics[width=0.9\linewidth]{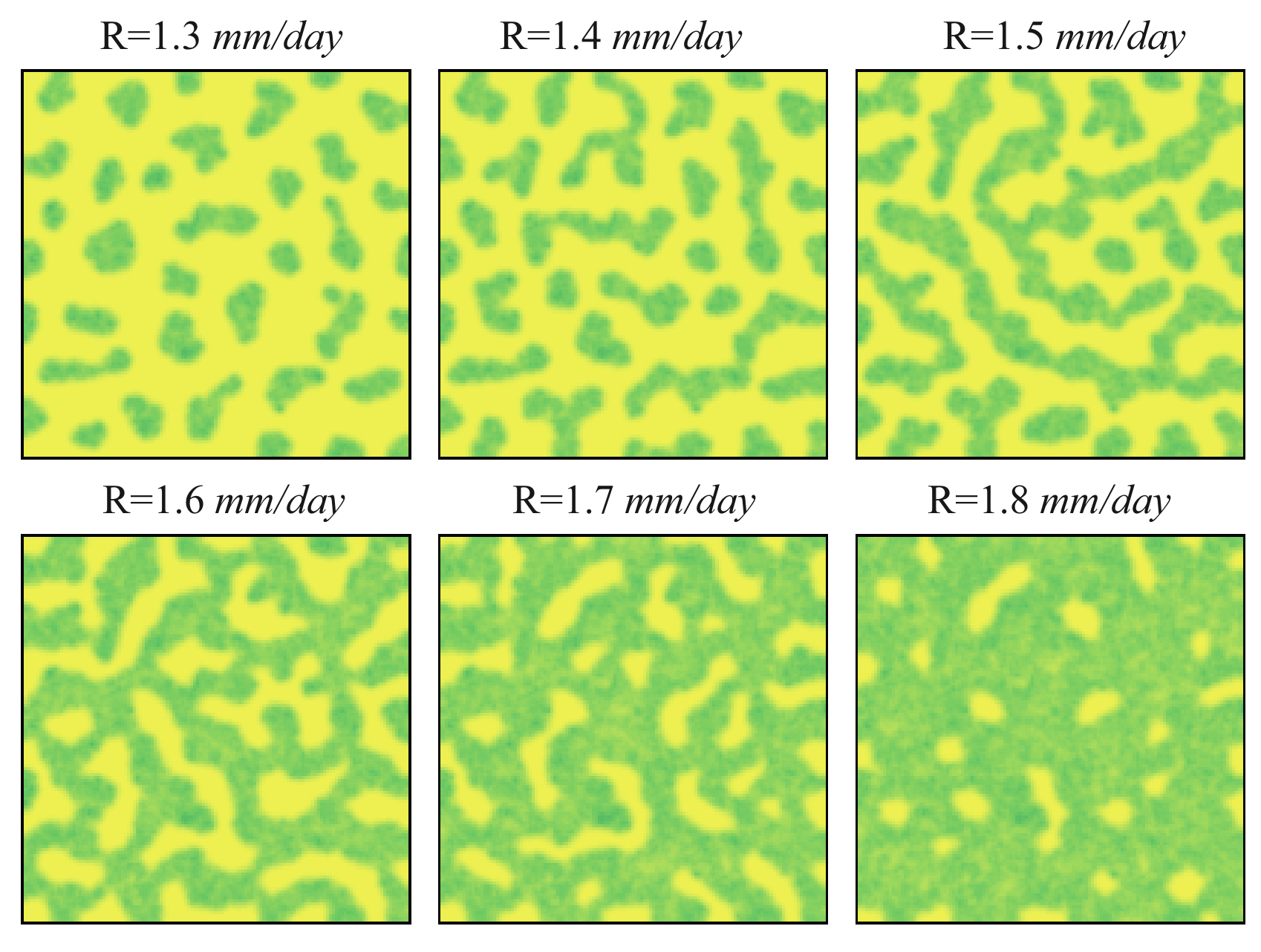}
   \caption{The basic states of a vegetation pattern in the presence of disorder. The patterns shown are the results of long-term simulations of the model (Eqs. \eqref{Rmodel}), and the different panels correspond to the precipitation rates specified. 
The average value of the evaporation and drainage rate is $\langle r_w \rangle=0.2/day$, and the standard deviation is $\sigma=0.02/day$. 
The bare-soil state at a low precipitation rate and the uniform vegetation cover at a high precipitation rate are not shown.}
   \label{fig:1}
 \end{figure}

To study the effect of quenched disorder, we used an extensively explored model for the dynamics of water-limited vegetation  \cite{HilleRisLambers2001ecology,Rietkerk2002a,rietkerk2004selfa,kefi2010bistability}. The model describes the spatio-temporal dynamics of three variables: the areal densities of vegetation, $B$, soil water, $W$ and surface water, $H$. Only one pattern-forming mechanism, the infiltration feedback, is captured by the model \cite{zelnik2013,kinast2014}. The infiltration feedback represents the increased infiltration rate of surface water in vegetation patches, thereby increasing the local vegetation growth rate. In bare-soil areas, a biological soil crust tends to form, reducing the infiltration rate. In addition, the plant root systems increase the porosity of soil and, thus, the infiltration rate. The total amount of water is conserved, and therefore, the increased soil-water density in the vegetation patches inhibits the growth in bare areas. The combination of local positive feedbacks and long-range competition for water leads to a finite wavenumber instability of the uniform state and to the pattern formation. 
The model equations are:
\begin{subequations}\label{Rmodel}
\begin{eqnarray}
 \frac{\partial B}{\partial t}&=&G_b B-\mu B+D_b\nabla^2 B; \\
 \frac{\partial W}{\partial t}&=&\mathcal{I}h-G_b B/c-r_w W+D_w\nabla^2 W; \\
 \frac{\partial H}{\partial t}&=&R-\mathcal{I}h-l_0H+D_h\nabla^2 H. 
\end{eqnarray}
\end{subequations}
The dynamics of the biomass density describes its soil-water-dependent growth rate, $G_b=cg\frac{W}{W+k_1}$ (this form ensures a linear dependency at a low soil-water density and saturation of the growth rate at a high soil-water density), the natural mortality rate, $\mu$, and its dispersal, by clonal growth or seed dispersion, which is characterized by the diffusion coefficient, $D_b$. The soil-water density grows due to surface-water infiltration whose rate depends on the biomass density, $\mathcal{I}=\alpha(B+k_2\phi)/(B+k_2)$. In patches of very high biomass density ($B>>k_2$), the infiltration rate is $\alpha$, while in bare-soil domains, it is reduced by a factor $\phi<1$ (the infiltration rate in bare-soil domains is $\alpha\phi$). The soil-water density is reduced by the vegetation water uptake, whose rate is $G_b/c$, and by the evaporation and drainage, whose rate is $r_w$. The spatial dynamics of the soil water consists of a simple diffusion with diffusivity $D_w$. It is well known that the diffusivity of soil water varies significantly with the soil-water content; however, for the temporal and spatial scales captured by the model, the description using a simple constant diffusivity is justified. 
\begin{figure*}[!ht]
   \centering
   \includegraphics[width=0.9\linewidth]{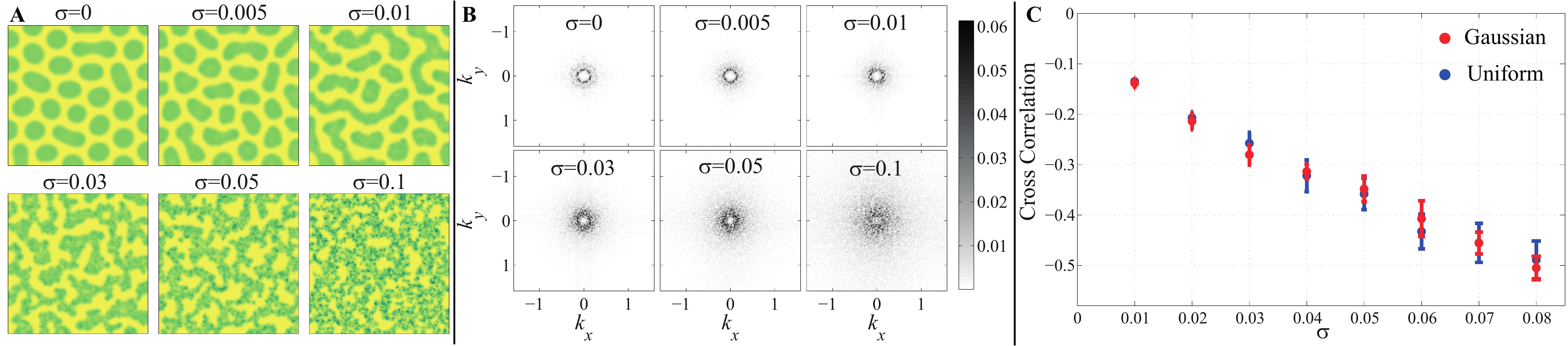}
   \caption{Panel A shows the effect of the disorder on the vegetation patterns. The different panels correspond to different degrees of heterogeneity (different values of the standard deviation of the evaporation and drainage rate, $\sigma$). All other parameters, including the mean value of the evaporation and drainage rate, are identical in all panels. The precipitation rate was set to $R=1.56 mm/day$, and all other parameters are specified in the text. In panel B, we show the Fourier amplitude of the patterns shown in panel A. For weak disorder, the self-organized pattern has a clear structure as seen by the clear peaks, while for strong disorder, there is no clear peak, and the pattern just follows the disorder. To complement the picture, we present, in panel C, the cross-correlation between the biomass and the disordered evaporation and drainage rate. For weak disorder, the cross-correlation is very weak, and the self-organization dominates the pattern formation, while for stronger disorder, there is a strong anti-correlation between $B$ and $r_w$, reflecting the fact that the biomass pattern is imposed by the disorder. The two datasets correspond to Gaussian and uniform distributions of the disorder. Clearly, the effect is independent of the exact distribution.}
   \label{fig:2}
 \end{figure*}
The surface-water dynamics accounts for the precipitation rate, $R$, the infiltration rate, $\mathcal{I}$, and the surface-water evaporation rate, $l_0$; for simplicity, we assume a plain topography in which the spatial dynamics of the surface water is described by simple diffusion with diffusivity $D_h$. The disorder was introduced in the evaporation and drainage rate, $r_w$. Variability in the soil texture, depth, composition and nutrient content can significantly affect the drainage rate \cite{mulla1999,sela2012soil} and motivates the use of this parameter as the disordered parameter. For simplicity, we considered two simple distributions of the disorder, a Gaussian distribution
\begin{equation}
 p_G(r_w)=\frac{1}{\sqrt{2\pi \sigma^2}}\exp{\left(-\frac{\left(r_w-\langle r_w\rangle\right)^2}{2\sigma^2}\right)},
\end{equation} 
which is fully characterized by its mean, $\langle r_w\rangle$ and variance, $\sigma^2$, and the uniform distribution (\eqref{udist}).
Different realizations of the quenched disorder are presented in the appendix \cite{SM}.
In what follows, we investigated the model in two spatial dimensions, and we used the following model parameters: $g=0.05[1/day]$, $k_1=5[kg/m^2]$, $\mu=0.25[1/day]$, $D_b=0.1[m^2/day]$, $\alpha=0.2[1/day]$, $\phi=0.2$, $k_2 =5[kg/m^2]$, $c=10$, $\langle r_w\rangle =0.2[1/day]$, $D_w=0.1[m^2/day]$, $l_0=0.06$,  $D_h=100[m^2/day]$. These values were adopted from previous studies \cite{kefi2010bistability}, to allow for an easy comparison with published results for homogeneous (non-disordered) systems.

The first expected effect of the disorder is a modification of the patterns formed by the vegetation (similar to the effect we saw in the Swift-Hohenberg model, Fig. \ref{fig:SHpatt}). We find that the same basic five states along a rainfall gradient appear in disordered systems. The steady state patterns under different precipitation rates are presented in Fig. \ref{fig:1}. In systems with no disorder, a state composed of domains of different stable states is unstable \cite{zelnik2013}, and one of the states will propagate and eventually cover the whole domain (note the difference between coexistence and bistability). The disorder allows for the coexistence of different patterns as a steady state (see the panels for $R=1.4,1.5$ in Fig. \ref{fig:1} and the panel for $\sigma=0.01$ in Fig. \ref{fig:2}A). This effect is important in explaining the observations of mixed patterns, such as those shown in the appendix \cite{SM}.

The effect on the pattern formed by the vegetation for different strengths of the disorder is shown in Fig. \ref{fig:2}A. Under the same precipitation rate, the vegetation formed different patterns due to the disorder. The pattern formation, in the presence of the disorder, is the result of the interplay between two effects, the self-organization and the imposed pattern. For weak disorder, the pattern is mostly determined by the self-organization interactions, and for strong disorder, the dominant effect is the disorder-imposed pattern. In Fig. \ref{fig:2}B, we show the Fourier transform amplitudes of the patterns shown in Fig. \ref{fig:2}A. The transition from a pattern with strong peaks, for the self-organized pattern, to a broad spectrum, for the disorder-imposed pattern, is easily seen. To further clarify the effect of the disorder, we show, in Fig. \ref{fig:2}C, the cross-correlation between $r_w$ and the biomass density, $B$. For weak disorder, there is very little correlation between the two. For stronger disorder, there is a significant anti-correlation, which reflects the fact that vegetation tends to grow in ``niches'' where the evaporation and drainage rate is small. The two datasets (red and blue) correspond to Gaussian and uniform distributions of the disorder, respectively. The error bars were derived from eight realizations of the quenched disorder for each of the distributions.
\begin{figure}[!ht]
   \centering
   \includegraphics[width=0.9\linewidth]{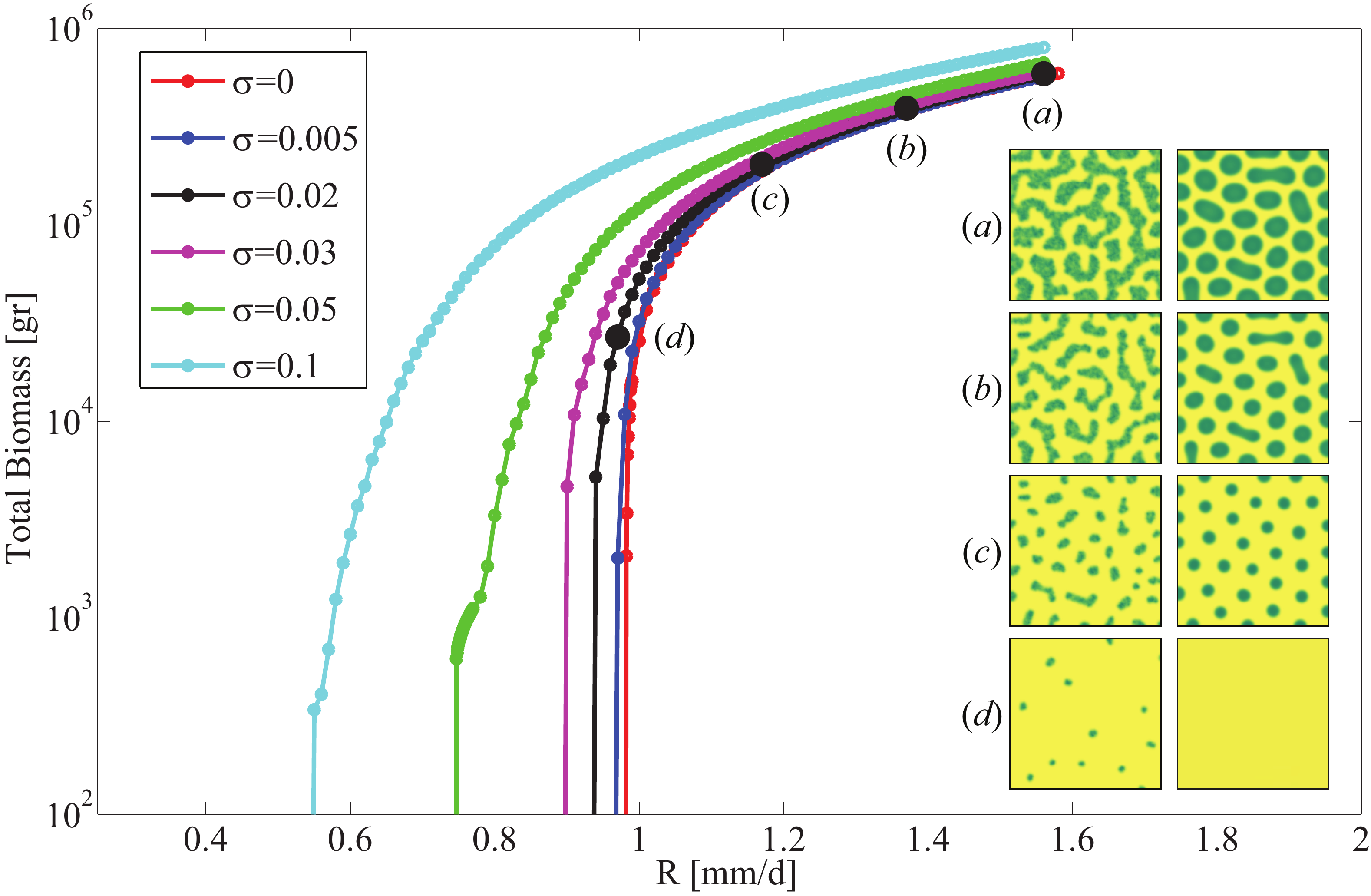}
   \caption{The effect of disorder on the durability of the system. The curves show the total biomass versus the precipitation rate for different strengths of the disorder. The more disordered the system, the lower the precipitation rate at which it collapses and the more gradual the transitions. The left panels in the inset show the vegetation patterns, in a system with $\sigma=0.02/day$, at precipitation values corresponding to the black dots ($R=1.56, 1.37, 1.17, 0.97$ $mm/day$ in panels (a), (b), (c) and (d), respectively). The panels in the right column of the inset show the vegetation patterns at the same precipitation rate values for the homogeneous system. Other model parameters are specified in the text.}
   \label{fig:3}
 \end{figure}

The most significant effect of the disorder is the increased durability. In Fig. \ref{fig:3}, we show the total biomass in the simulated system versus the precipitation rate. The more disordered the system (the larger $\sigma$ is), the more deleterious conditions (lower precipitation rate) it can survive. In addition, the transition from the vegetation pattern to the bare-soil state becomes more gradual as the disorder increases. The inset panels of Fig. \ref{fig:3} show the vegetation patterns at precipitation values corresponding to the dots. The left column of the inset shows the patterns for the disordered system, $\sigma=0.02$, and the right column shows the patterns for the homogeneous system. The inset panels demonstrate the fact that the vegetation in the disordered system survives under conditions in which the homogeneous system collapses to the bare-soil state.

\begin{figure}[!ht]
   \centering
   \includegraphics[width=0.9\linewidth]{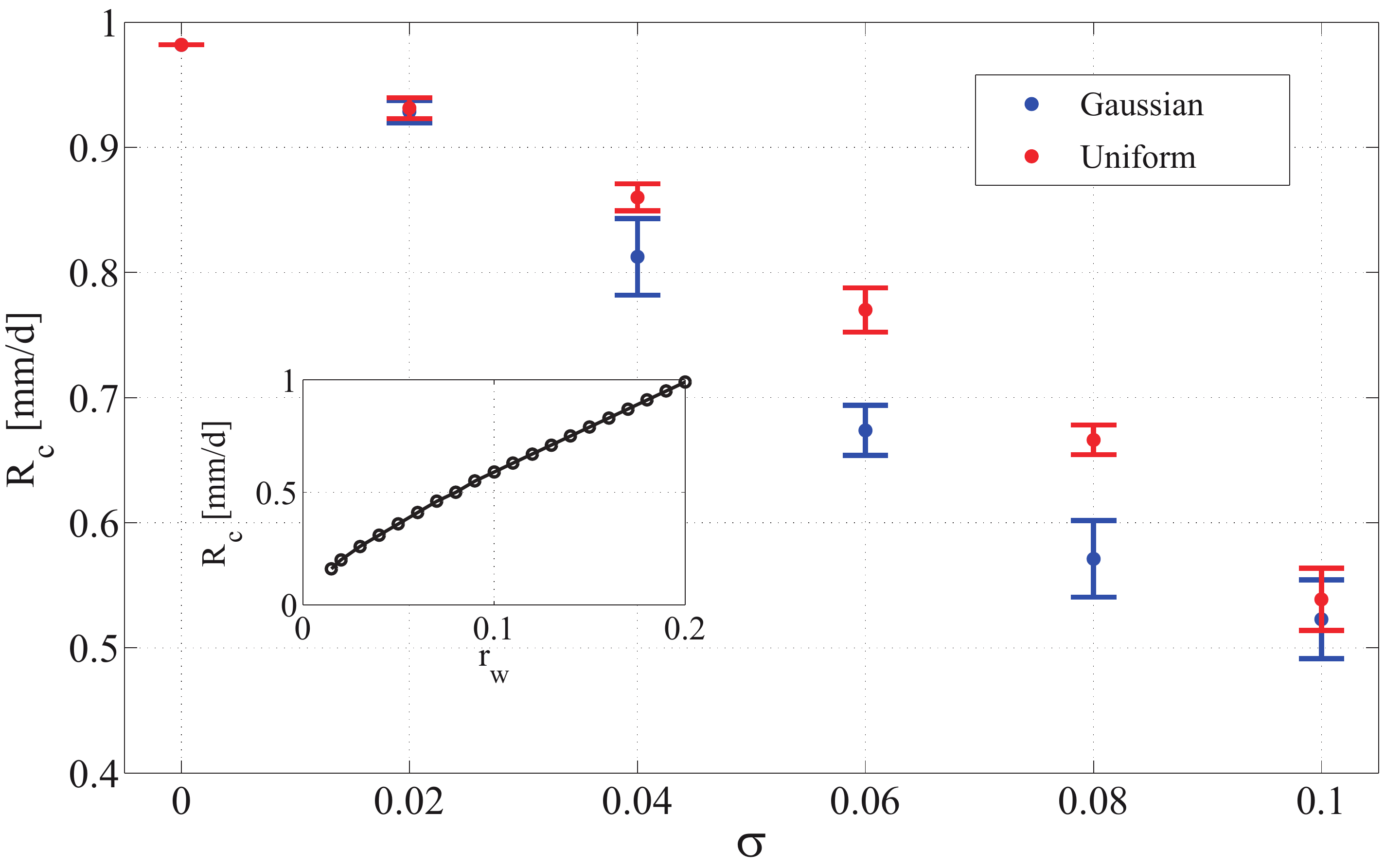}
   \caption{The critical precipitation rate, $R_c$, versus the standard deviation of $r_w$'s probability density function. The error bars represent the standard deviation in the
value of $R_c$ as derived from eight realizations of the quenched disorder. The inset shows $R_c$ for the case of no disorder and different values of $r_w$. The two datasets correspond to Gaussian and uniform distributions of the disorder.}
   \label{fig:4}
 \end{figure}

In Figure \ref{fig:4}, we present the critical value of the bifurcation parameter, $R_c$, for which the vegetation collapses into the bare-soil state ($B=0$). The two datasets (blue and red) correspond to Gaussian and uniform distributions of the disorder, respectively. The error bars were derived from eight realizations of the disorder for each value of $\sigma$ and each distribution. The qualitative behavior, namely the increased durability for stronger disorder, is the same for both distributions. The inset shows $R_c$ versus $r_w$ for homogeneous systems. 

\begin{figure}[!ht]
   \centering
   \includegraphics[width=0.9\linewidth]{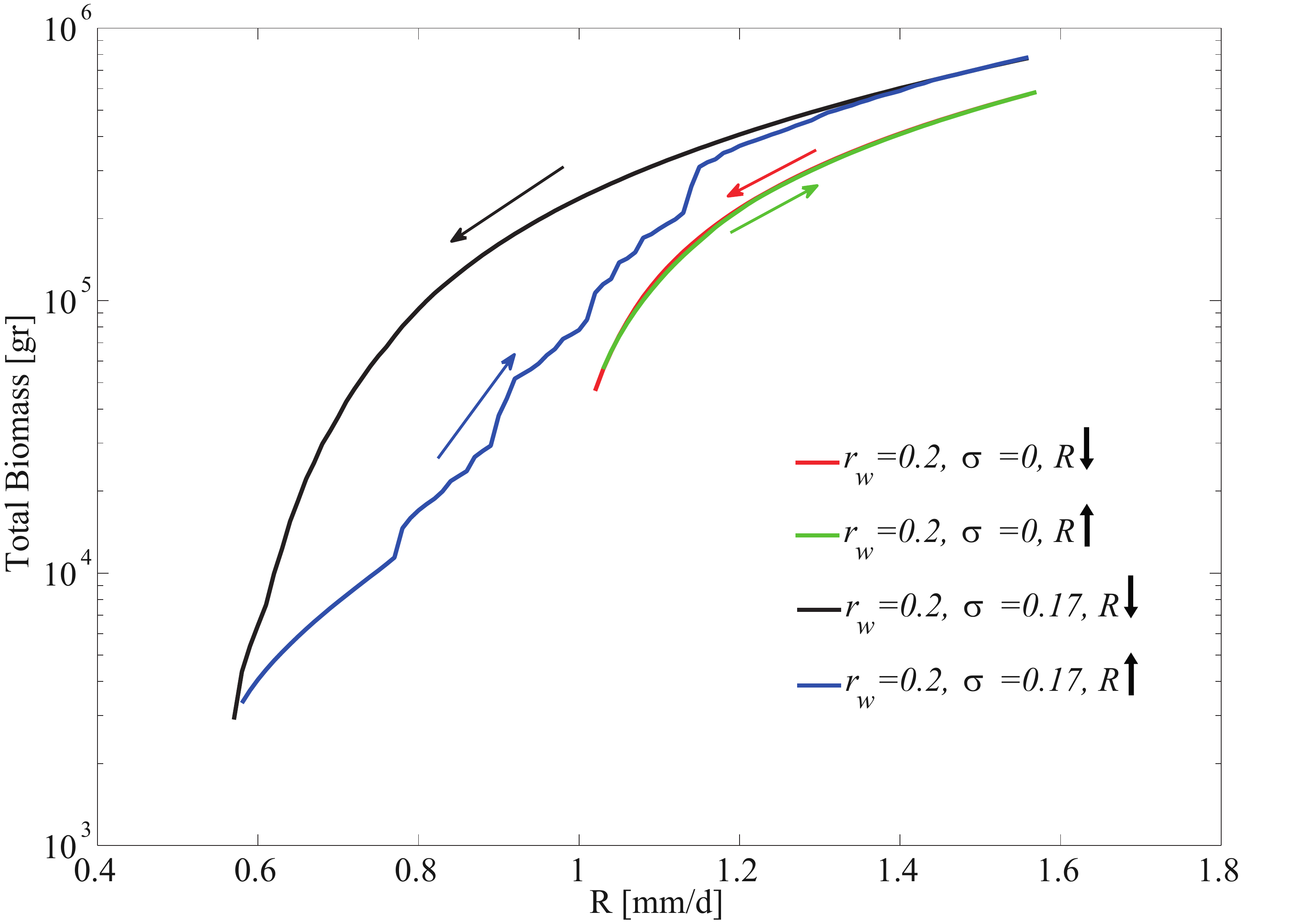}
   \caption{Heterogeneity-induced hysteresis. The black line shows the total biomass versus the precipitation rate as it decreases. The reduction of
the precipitation rate is stopped before the system collapses. The blue line shows the total biomass (in grams) versus the precipitation rate as it increases from this
point. Clearly, the system does not recover along the same path but follows a different path with a lower vegetation biomass. To demonstrate that
this is a heterogeneity effect, we present the red and green lines showing the same quantities for a system with no heterogeneity. All the model
parameters are the same as those specified in the text.}
   \label{fig:hysteresis}
 \end{figure}

Another important effect of the disorder is seen when investigating the resilience of the system. When the precipitation rate is decreased, the total biomass is decreased and vice versa.
In the absence of disorder, the dynamics of the system in both directions, increase and decrease of the bifurcation parameter, occurs along the same line in the bifurcation diagram as shown by the red and green lines in Figure \ref{fig:hysteresis}. However, the disorder induces a hysteresis in the response of the system as shown by the black and blue lines in Figure \ref{fig:hysteresis}. It is important to note that this type of hysteresis occurs well before the critical transition from one basic state to another.

\section{Discussion and Summary}\label{sec:disc}
Our results show that as expected, the quenched disorder affects the patterns formed. In the Swift-Hohenberg model, the modification appears mainly as distortions of the periodic patterns. Interestingly, the interplay between the self-organized and imposed patterns not only modifies the pattern formed but also allows the coexistence of several patterns, such as labyrinthine and spotted patterns, as a stable state of the system. The latter is easily seen in the results for the model for dryland vegetation dynamics, which was simulated in two spatial dimensions that allowed patterns with different symmetries. 

We showed that as the disorder increases, the pattern formed shifts from the self-organized pattern to the imposed pattern. The disorder allows for different domains to be in different states, thereby not only affecting the patterns but also increasing the durability of the system and making the transition to the bare-soil state more gradual. The cross-correlations between the disordered parameter and the dynamical variable reveal that there is an essential difference between pattern-forming systems and systems exhibiting a bistability of uniform states. In the latter type of systems, the cross-correlations are very high for all values of the bifurcation parameter, reflecting the fact that the dynamical variable responds locally to better/worse conditions. However, for pattern-forming systems, the cross-correlations are lower due to the short-range positive feedbacks that allow the dynamical variable to respond to the conditions in the vicinity (and not just to the local conditions). Interestingly, this behavior suggests that the cross-correlations may serve as an indicator for approaching critical transitions. As the system approaches the critical transition, the cross-correlations are stronger due to the fact that the dynamical variable survives only in preferred niches. Similar, but much weaker, results were obtained for disorder in the soil-water diffusivity \cite{yizhaq2014}. The reason for the weaker effect of disordered diffusivity is the spatial averaging by the nonlocal term--the spatial derivatives. The averaging narrows the effective width of the diffusivity distribution, thereby reducing the effects of its disorder.

Interestingly, the disorder qualitatively affects the dynamics of the system even far from the critical point. In pattern-forming systems, the feedbacks representing the short-range facilitation can stop the propagation of fronts between domains in different states, and therefore, the disorder strongly affects the dynamics of the system. We showed that the disorder induces a hysteresis in the response of the system to adiabatic changes in the bifurcation parameter. For homogeneous systems, the dynamics is reversible, while in disordered systems, local domains shift from one state to another and only recover at higher values of the bifurcation parameter, thereby inducing the hysteresis. 

In addition to the fundamental Swift-Hohenberg model and the model for dryland vegetation dynamics, we also used other pattern-forming models and found qualitatively the same results as those presented here. Therefore, we believe that the results are valid for any pattern-forming system. Spatially extended systems (and, in particular, ecosystems) are often heterogeneous, and therefore, the effects of quenched disorder cannot be ignored when studying critical transitions and, in particular, early warning signals for imminent regime shifts. 
A very interesting question is how spatially correlated disorder affects the dynamics and bifurcation diagrams of nonlinear systems. It is expected that the interplay between the correlation length of the disorder and the typical length of the self-organized pattern will also affect the system. The significance and relevance of the results presented here extend beyond the context of vegetation dynamics; the findings emphasize the importance of studying the effects of quenched disorder in a wide array of physical systems, including condensed matter (in particular, vortex matter in type-II superconductors, which is strongly affected by quenched disorder \cite{fisher1991}), nonlinear optics, chemical reactions, sand dune dynamics and many others.

\begin{acknowledgments}
The research leading to these results has received funding from the European Union Seventh
Framework Programme (FP7/2007-2013) under grant number 293825.
\end{acknowledgments}

\appendix*
\section{}

\begin{figure}[!ht]
   \centering
   \includegraphics[width=0.9\linewidth]{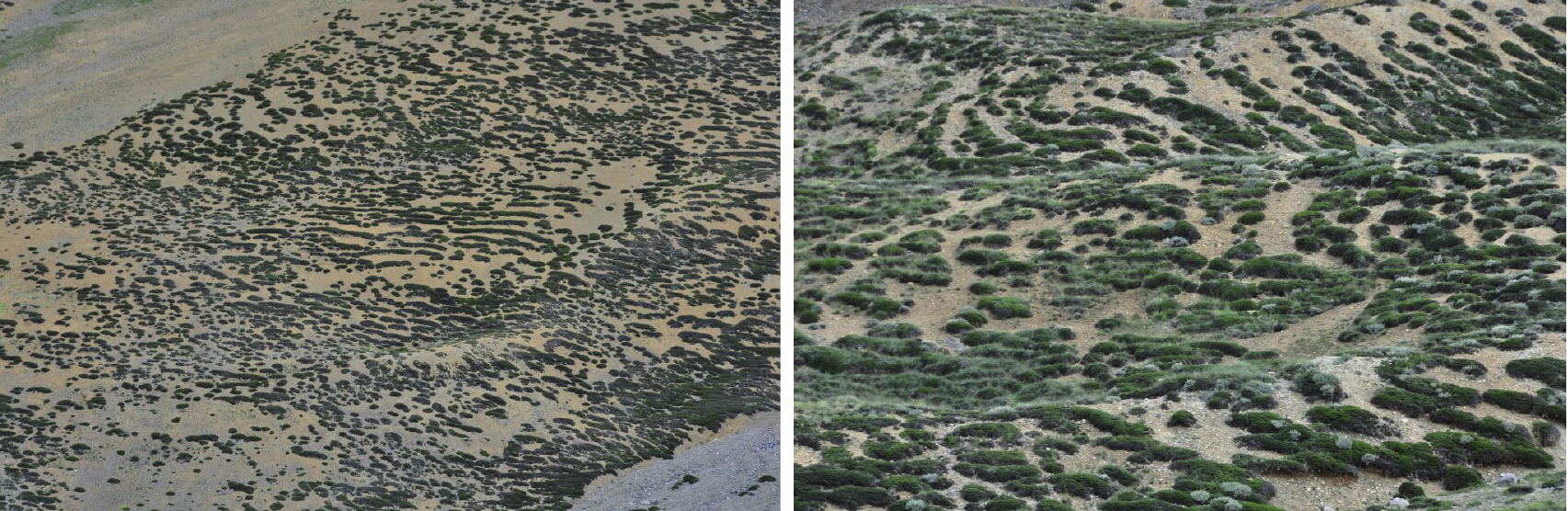}
   \caption{Vegetation patterns in Ladakh, India ($32^{\circ}$ $53^{\prime}$ $24^{\prime\prime}$ N,$77^{\circ}$ $31^{\prime}$ $48^{\prime\prime}$ E). The altitude is $\sim 4500 \hskip 3pt m$ above sea level.
 The mean precipitation rate in this area is less than $200\hskip 3pt mm/yr$. The coexistence of labyrinthine/striped and spotted patterns is remarkable.}
   \label{fig:1SM}
 \end{figure}
\begin{figure}[!ht]
   \centering
   \includegraphics[width=0.7\linewidth]{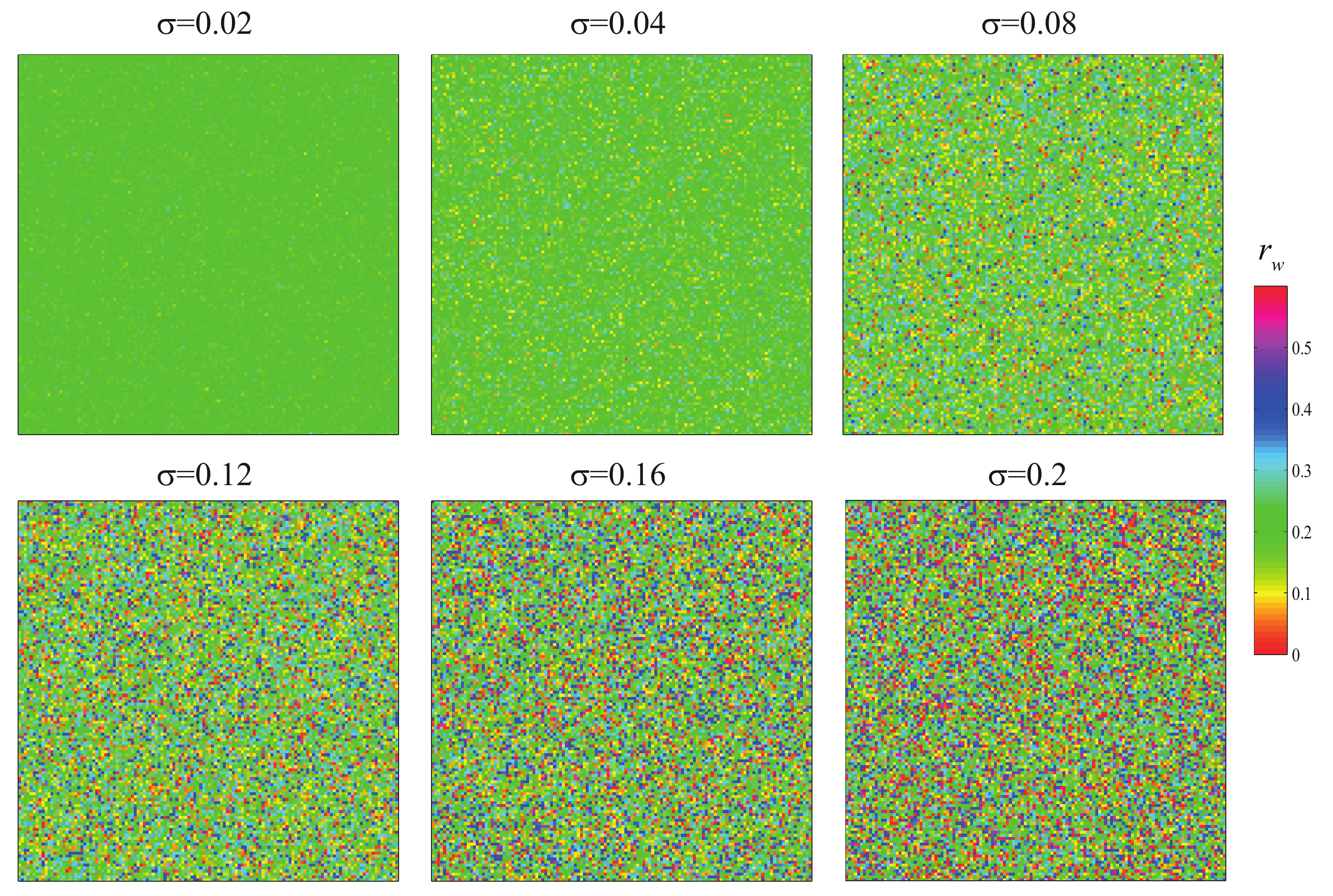}
   \caption{Realizations of the spatial distribution of the evaporation and drainage rate, $r_w$, for
different values of $\sigma$. In all panels, $\langle r_w \rangle=0.2/day$. The realizations were derived from the Gaussian distribution. For the uniform distribution, the realizations look qualitatively the same.}
   \label{fig:2SM}
 \end{figure}
\begin{figure}[!ht]
   \centering
   \includegraphics[width=0.5\linewidth]{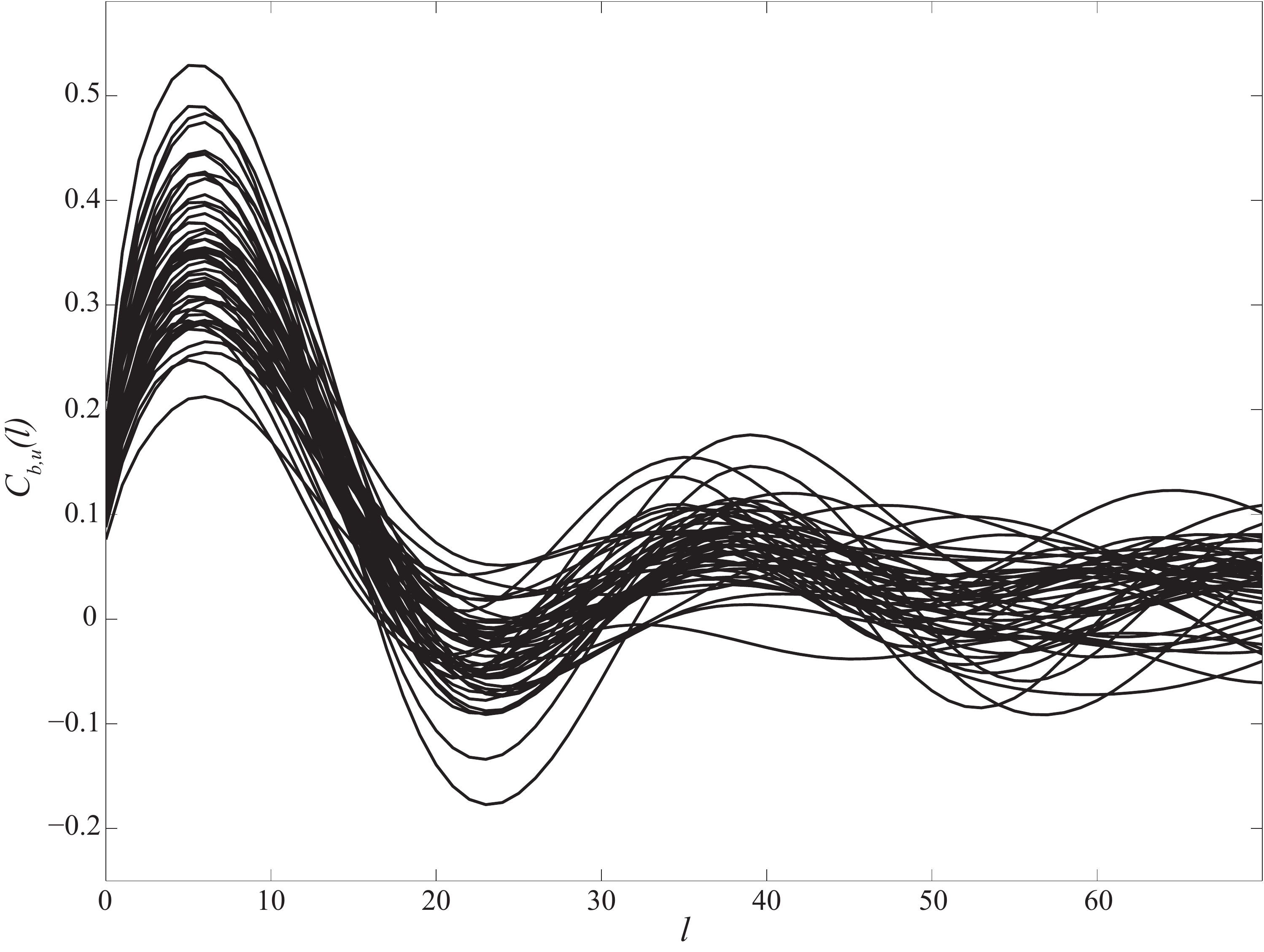}
   \caption{The cross-correlation between the disordered parameter, $b$, and the dynamical variable, $u$, as a function of the size of the domain ($l$) over which the value of $b$ is averaged. The different lines correspond to different realizations of the disorder. The cross-correlations were calculated for the Swift-Hohenberg model and the parameters specified in the main text. The bifurcation parameter, $r=0.92$. $b$ was drawn from a uniform distribution with width $v=2$.}
   \label{fig:CCl}
 \end{figure}

The cross-correlation between the disordered parameter and the dynamical variable is suggested as an early indicator for imminent regime shifts. 
We mentioned in the main text that due to nonlocal terms, the cross-correlation is maximal when the average value of the disordered parameter in the vicinity of each point is considered rather than the local value.
The size of the domain is determined by the nonlocal term whose length scale is dictated by the parameter $q_c$. In order to find the optimal size of the domain, we calculated the following cross-correlation function:
\begin{equation}
 C_{b,u}(l)\equiv \frac{\frac{1}{N}\displaystyle\sum\limits_{i=1}^N\left(\left(\sum\limits_{i-l}^{i+l}b\left(x_i\right)\right)-\langle b \rangle\right)\left(u\left(x_i\right)-\langle u \rangle\right)}{\sqrt{\displaystyle\sum\limits_{i=1}^N\left(u\left(x_i\right)-\langle u \rangle\right)^2}\sqrt{\displaystyle\sum\limits_{i=1}^N\left(\left(\sum\limits_{i-l}^{i+l}b\left(x_i\right)\right)-\langle b \rangle\right)^2}}.
\end{equation}
Here, $b$ is the disordered parameter, $u$ is the dynamical variable, $N$ is the number of sites simulated, $l$ is the length of the domain over which the disordered parameter is averaged and the angular brackets denote spatial averaging over the whole simulated domain. In Figure \ref{fig:CCl}, we present the cross-correlation versus $l$ for different realizations of the disorder. In this figure, we used the Swift-Hohenberg model with the parameters specified in the main text and $r=0.92$. The parameter $b$ was drawn from a uniform distribution of width $v=2$. One can easily see that while the maximal cross-correlation varies between the different realizations of the disorder, it is obtained for the same size of the spatial averaging domain, $l\approx5-7$ (because the size of each grid cell is $dx=65\pi/1024\approx0.2$, it implies a domain of size $\approx 1$). We verified that the size of the domain yielding the maximal cross-correlation is independent of the distribution of the disordered parameter and also independent of the bifurcation parameter, $r$. In our dimensionless model, it is dictated by the length scale of $1/q_c=1$. The results presented in Figure 3 of the main text were derived using $l=6$.

\clearpage

%
\end{document}